# Lessons from DEPLOYment

Manuel Mazzara, Cliff Jones, Alexei Iliasov

School of Computing Science, Newcastle University, UK

## 1    Overview

This paper reviews the major lessons learnt during two significant pilot projects by Bosch Research during the DEPLOY project [1]. Principally, the use of a single formalism, even when it comes together with a rigorous refinement methodology like Event-B, cannot offer a complete solution. Unfortunately (but not unexpectedly), we cannot offer a panacea to cover every phase from requirements to code; in fact any specific formalism or language (or tool) should be used only where and when it is really suitable and not necessarily (and somehow forcibly) over the entire lifecycle.

## 2    DEPLOY and its scenarios

DEPLOY is an ambitious project addressing diverse major industrial areas: automotive, train transportation, business and aerospace software. Putting all of them under the same hat is a difficult (if not impossible) task for both instrinsic and extrinsic reasons. We recognised from the very beginning that each deployment scenario would be different and would need (at least partially) different approaches, concepts and tools. Industries are different, development process differs, internal organizations are varied and, to some extent, business models are different as well as politics. Most importantly, target applications and inhouse engineering tools/standards show little similarity. Therefore, integration needs are different.

    The RODIN project [2] constituted a solid basis for DEPLOY giving us valid reasons to claim that the Rodin tools [3] were moderately mature. The tools were well designed by an outstanding team and the experience, including with industrial partners (although RODIN was a STREP rather than an IP), taught us that tool support is essential for most technology transfer undertakings. Although at the end of RODIN we were aware of the need to support the tools and their evolution, the reality we had to face in DEPLOY was tougher than expected for both academia and industry. Effort was immediately put in place to train industry partners via a "block course" to provide knowledge and skills leading to mini-pilots

first, and then major case studies (in the case of Bosch specifically two: Cruise Control and Start/Stop system [4]).

## 3   Problems and opportunities

The Event-B notation favours a style of modelling that may be loosely characterised as event-based or reactive. There are examples where this works well and models appear natural and elegant. Unsuprisingly, there are also cases where there is a misfit between the style and developers' expectations or system requirements. This issue is more pronounced where an industrial user is involved. In an academic setting, there is a greater degree of flexibility of how a model is developed since the construction of a piece of software is rarely an end in itself. A researcher has the advantage of a broader perspective giving the ability to recognise, without investing considerable effort, that a method is not suitable for a given problem. In the DEPLOY project, few industrial users had prior exposure to formal methods and their perspective was limited to Event-B method and tools. Obviously, they also did not have an opportunity nor the desire to avoid, or adapt their problems to the Event-B style. This steadfastness, while at times frustrating for academic partners, has resulted in a stream of feature requests, some of which were, at least partially, addressed by offering methodological and tooling extensions.

Once beyond the stage of the mini-pilots, all the industrial users raised the issue of notation expressiviness. Many concepts found in programming and specification languages (records, procedure calls, macro definitions, modules, polymorphic types, meta-theorems, etc.) are not part of core Event-B. This is not a fundamental problem since the Rodin Platform is designed to be extensible and the core Event-B was purposely made compact. Still, the tool developers were surprised when industrial users indicated what they believed to be essential features missing in the Platform. The reaction took some time as effort had to be reallocated from already planned activities. In the meanwhile, industrial users still proceeded without these additional tools.

At least four tools were developed in response to Bosch team requests (group refinement, records, flow and team work) and some methodological work was inspired by the problems the Bosch team has encountered applying Event-B. With hindsight, the tool developers should have spent more time with the industrial partners compiling tool requirements. Initial version of these additional tools did not fit the industry expectations.

The transition from mini-pilots to larger case studies has identified a number of weaknesses in the tool implementation. The user interface did not cope well with machines comprising even a few dozens of events; this was further aggravated by hard to reproduce resource handling bugs. Such problems would, perhaps, not have been taken seriously for a long time were there not some commited users who constructed the first large-scale Event-B models. For the Bosch team, the main weak points of the tool turned out to be text editing; slow interactive prover UI; and liveness proof obligations in the form of gigantic disjunctions. The text editor, a suprisingly involved tool due to the Platform design, has improved immensily in the past year while the prover UI has been redesigned using a different rendering technology. There is no easy solution for proving complex liveness proof obligations but, in many cases, the SMT solving facility of the ProB plug-in has had spectacular success.

These problems lead us to note a contrast between models from academia and industry. Research models are rarely large (and need not be for the focus is on challenging aspects that are best studied in isolation). On the other hand, an industrial development aims at a product and works from a real requirements document. This results in a large model which is often fairly "shallow" in the sense that many parts are merely descriptive and do not entail deep verification properties. It turns out that large models require a different treatment. As one example, in a terse, academic model, a message of a communication protocol would be some value $m \in M$. Message contents, like destination address or payload, could be added when necessary by defining mappings from $M$. But an industrial user deals with a detailed description of protocols where messages have dozens of attributes. Not only it is harder to find good abstraction and then carry out many (trivial but cumbersome) data refinement steps but the end result is far from elegant as a large number of variables are necessary to define what is conceptually a single entity. This specific issue has been addressed by the records plug-in but one can find many similar concerns stemming from the same problem of method, notation and tool scalability.

To make a development viable, a large model must be split at some point into sub-models. There are technical solutions for this but industrial partners have found them difficult to use. At the moment, formal model decomposition remains an art that requires skill and experience. To develop a real-life product, it is essential to be able to share the development effort among a team of modellers (this is partially addressed by the team work plug-in) and reuse existing modelling artefacts (to a limited extent addressed by the three decomposition techniques).

There are methodological issues arising from attempting large scale developments. It is not known how to plan a refinement strategy when faced with a detailed requirements document. It is impossible to foresee all the major design decisions which turn development into a trial and error. This itself would not be such a problem were it not so difficult and expensive (in terms of lost proofs) to refactor refinement chains in Rodin.

## 4   Achievements and Lessons

The intense collaboration with Bosch Research has been a valuable learning experience for both sides. We had the chance to approach several software engineering issues, contributing to some and, unavoidably, leaving others open. We believe our work has clarifed several aspects of industrial deployment of formal methods in automotive applications. Most important of all, we realized the limitations of Event-B, both as a formalism and as a method. The lack of a rigorous and repeatable approach of many other "formal methods" is well known. In [5] and [6] this issue is historically investigated and the requirements over a "formal method" are identified to discover that many methods are actually just notations, i.e. just formalisms without an attached rigorously defined and repeatable, systematic approach. Event-B is *not* one of those. Its refinement strategy was demonstrated to be useful when applied to several case studies in a number of projects like RODIN and DEPLOY. However, not even Event-B is a panacea applicable to every phase of software development. Instead of attacking every problem with a single weapon, we opted to use a portfolio of different instruments, which is an idea also supported by other researchers ([7], [8]). The overall strategy was demonstrated to be successful and, given the thorough documentation generated by the project ([9], [10], [4]), it promises to be repeatable by engineers with an initially limited knowledge of formal methods. However, the role of training cannot be underestimated and limitations of the current "knowledge transfer" approach have been identified. For example, the block course organized to train industry partners in 2008 was a valid choice, but more specific industry needs have emerged showing how closer and prolonged interactions are generally preferable. The documentation available at the beginning of DEPLOY had weaknesses, leading to the decision to generate a new user manual [11]. The importance of well written, high quality and complete documentation can never be emphasized enough (and Bosch, in particular, raised this point since the very beginning). The actual deployment consisted in formal modelling of two major relevant applications for Bosch: the Cruise Control and the Start/Stop system. Two different methodologies have been applied to the case studies as described in detail in [4], which also describes the motivations behind the choices. This process gave us a much better understanding of the links between requirements and Problem Frames and, in turn, the relationships with Event-B models.

Several lessons have been learnt during this intense and exciting experience. First of all, we had confirmation of other researchers' experience, i.e. the use of a single formalism cannot offer a complete solution to large problems. The primary importance of tool support stands not only in its mere existence, but in its ability to meet specific industry needs. Usability and performance, for example, have not been considered sufficiently before deployment and therefore became critical aspects in the process. Having tool support without users being able to actually use it is of little consolation. However, during the project several industry requirements have been satisfied thanks to the good feedback received. Finally, documentation and training are major aspects of deployment and crash courses do not seem to offer a solution. This is why "education strategies" should be implemented differently in future projects.